\documentclass[iop,revtex4]{emulateapj}
\usepackage{lineno}

\begin{document}

\renewcommand{\topfraction}{1.0}
\renewcommand{\bottomfraction}{1.0}
\renewcommand{\textfraction}{0.0}

\title{Orbit alignment in triple stars}

\author{Andrei Tokovinin}
\affil{Cerro Tololo Inter-American Observatory, Casilla 603, La Serena, Chile}
\email{atokovinin@ctio.noao.edu}

\begin{abstract}
Statistics of  the angle $\Phi$ between orbital  angular momenta in hierarchical
triple  systems with  known  inner visual  or  astrometric orbits  are
studied.   Correlation between  apparent revolution  directions proves
  partial  orbit  alignment  known  from   earlier  works.  The
alignment is  strong in triples  with outer projected  separation less
than $\sim$50  AU, where  the average $\Phi$  is about  $20\degr$.  In
contrast, outer  orbits wider  than 1000 AU  are not aligned  with the
inner orbits.   It is established  that the orbit  alignment decreases
with increasing  mass of the primary  component.  Average eccentricity
of inner  orbits in well-aligned  triples is smaller than  in randomly
aligned  ones.   These  findings  highlight the  role  of  dissipative
interactions with gas in defining the orbital architecture of low-mass
triple systems.  On the other hand, chaotic dynamics apparently played
a  role in shaping  more massive  hierarchies.  Analysis  of projected
configurations and triples with known inner and outer orbits indicates
that  the distribution  of $\Phi$  is  likely bimodal,  where 80\%  of
triples  have $\Phi  < 70\degr$  and the  remaining ones  are randomly
aligned.
\end{abstract} 

\maketitle

\section{Introduction}
\label{sec:intro}

Statistics of the angle $\Phi$  between orbital angular momenta of the
inner  and outer  orbits in  triple systems,  considered  jointly with
other parameters such  as eccentricity and mass, can  inform us on the
relative  role   of  dynamical  and  dissipative   processes  in  star
formation.   Orientation of angular  momentum of  stars, circumstellar
disks,  and planets  is determined  by the  same processes,  hence the
study of hierarchical stellar systems is relevant to these problems.

Some hierarchical   systems in  the field have  misaligned and
even  counter-rotating  inner  and  outer orbits,  e.g.   $\sigma$~Ori
\citep{Schaefer2016}   or  $\zeta$~Aqr   \citep{ZetaAqr}.    The  {\it
  misaligned}  triple systems  could have  been shaped  by dynamical
interactions   between  stars   in  clusters   or  in   small  groups.
\citet{Antognini2016}  show that typical  triple stars  resulting from
dynamical  interactions   have  small  ratios of  outer  to  inner
separation  and random  relative  orbit orientation.   Both inner  and
outer  eccentricities in such  hierarchies have  approximately thermal
distributions $f(e) = 2e$, favoring eccentric orbits.

On the  other hand, there are  many hierarchies where  the orbits have
modest mutual inclinations and eccentricities \citep{Planetary,Tok17}.
Those  multiple systems  with planetary-like  architecture  (call them
{\it planar}  for brevity) presumably  formed or evolved in  a gaseous
disk.  The role  of dissipation in aligning multiple  systems has been
demonstrated    by   the    large   hydrodynamical    simulations   of
\citet{Bate2014}.   He  found  $\langle  \Phi \rangle  =  39\degr  \pm
7\degr$, with  closer triples being better aligned  (see his Fig.~20).
It  is not  known  presently  whether there  are  indeed two  distinct
families of  multiple stars formed  by these alternative  scenaria, or
if the misaligned and planar triples represent two extremes in one common
population.

The  angle $\Phi$  between inner  and outer  angular  momenta (orbital
spins)  can  vary  between  0  and 180\degr.   For  uncorrelated  spin
directions, $\cos \Phi$ is distributed uniformly between $-1$ and $1$,
hence $f(\Phi) \propto \sin \Phi$,  making $\Phi \approx 90 \degr$ the
most likely angle.  In such case, 78\% of triples have $39\degr < \Phi
<  141\degr$ and  their  relative inclination  and inner  eccentricity
change   periodically  in  the   Kozai-Lidov  cycles   \citep[see  the
  references  in][]{Antognini2016}.    In  these  cycles,   the  inner
eccentricity can reach high values.  When two stars on eccentric orbit
approach each other within a few stellar radii, tidal forces come into
play  and   shorten  the  inner   period,  creating  a   close  binary
\citep{KCTF}.     Statistical    study    of   this    mechanism    by
\citet{Fabrycky2007}  assumed initial  triples  with randomly  aligned
orbits;  the  final  inclinations  concentrate  around  $\Phi  \approx
40\degr$  ~and  $\Phi  \approx  140\degr$,  i.e.  half  of  the  close
binaries are counter-rotating, while the frequency of $\Phi \sim 90\degr$ is
reduced.  In reality,  counter-rotation is rare \citep{Borkovits2016},
implying that the  initial distribution of $\Phi$ was  not random.  If
triple   systems   are   approximately   aligned  at   birth,   as   in
\citep{Bate2014}, formation  of close binaries by  the Kozai mechanism
becomes much less frequent than surmized by \citet{Fabrycky2007}.

The angle  $\Phi$ can be measured  directly only in a  small number of
resolved  hierarchical  systems.  The  statistics  of  $\Phi$ is  best
studied by  indirect techniques  relating some observable  quantity to
the distribution of  $\Phi$.  The general trend of  orbit alignment in
triple  systems   has  been  demonstrated   for  the  first   time  by
\citet{Worley1967}.   He compared  the numbers  of apparently  co- and
counter-rotating  visual   triples  and  interpreted   their  relative
frequency in terms of the  average angle $\langle \Phi \rangle$, which
he found to  be about $50\degr$.  This simple  yet powerful method has
been used in the following  studies of orbit alignment, including this
one.   \citet{Tok93} applied  several statistical  methods  to
different  subsets of  triple systems.   His results  were  updated by
\citet{ST02} who attempted to  match the observed partial alignment by
dynamical simulations of small decaying clusters.

The present study takes advantage  of the increased number of multiple
systems with  known orbits,  allowing us for  the first time  to probe
orbit  alignment as  a function  of component's  mass  and separation.
Indeed,  considering all  triple stars  as one  population is  a crude
assumption adopted in  earlier works out of necessity,  because of the
samll samples; it can be  dropped now.  The sign correlation pioneered
by Worley  is the main method used  here (Section~\ref{sec:sign}).  It
is supplemented  by the study of  triple systems with  known outer and
inner orbits (Section~\ref{sec:vbvb}) and  by the analysis of apparent
configurations of triples (Section~\ref{sec:config}).  The results are
discussed  in Section~\ref{sec:disc}  and compared  to the  studies of
disk  alignment  in  young  binaries   and  to  the  recent  work  by
\citet{Borkovits2016} on compact triples.

\section{Data}
\label{sec:data}

The data on hierarchical systems  are extracted from the Multiple Star
Catalog,    MSC    \citep{MSC}.     Its    latest    (2010)    on-line
version\footnote{\url{http://www.ctio.noao.edu/\~{}atokovin/stars/index.php}}
has  been augmented  by adding  new  multiples from  the 67-pc  sample
\citep{FG67}, results  of speckle interferometry \citep[e.g.][]{SOAR},
and the  literature.  The latest version of the  visual orbit
catalog  VB6   \citep{VB6}  was  queried  to  add   new  orbits.   The
preliminary  version of  the updated  MSC is  posted online.\footnote{
  \url {http://www.ctio.noao.edu/\~{}atokovin/newmsc.tgz}}

We  extracted  from  the  MSC  triple systems  with  known  visual  or
astrometric  inner  orbits (types  V  or  A)  and a  distant  tertiary
component.  For the purpose of  this study, we ignore additional outer
and  inner  subsystems,  i.e.   consider higher-order  hierarchies  as
simple ``triples''.    The sample contains 274 triples and  138 systems  of four or
  more stars.  Each of  the 19 2+2  quadruples with both  inner orbits
  known is listed as two triples with the same outer separation.

\begin{deluxetable*}{l rr  rccrr  rr rr c }[ht]                                                                                                                                
\tabletypesize{\scriptsize}                                                                                                                                                     
\tablecaption{Hierarchical systems with resolved inner orbit (fragment)                                                                                                                             
\label{tab:cpmvb} }                                                                                                                                                            
\tablewidth{0pt}                                                                                                                                                                
\tablehead{ \colhead{WDS} & \colhead{$\pi_{\rm HIP}$} &  \colhead{$M_1$} 
& \colhead{$P_1$} & \colhead{$e_1$} & \colhead{$a_1$} & \colhead{$\Omega_1$ } & \colhead{$i_1$} & 
\colhead{$\rho$} &   \colhead{$\theta$} & \colhead{Revolution} & \colhead{Sign}  \\
& \colhead{(mas)} &  \colhead{(${\cal M}_\odot$)} & 
\colhead{(yr)} &  &  \colhead{(\arcsec)} & \colhead{(\degr)} &  \colhead{(\degr)} &
 \colhead{(\arcsec)} & \colhead{(\degr)} &\colhead{Direction}  & 
   }   
\startdata     
00024+1047 &   11.4 &   1.04 &  129.7 &   0.046 &  0.366 &   59.4 &   97.2 &   63.20 &  301.0 &  0 &  0 \\
00046+4206 &    6.7 &   3.05 &   70.1 &   0.515 &  0.165 &  100.6 &  104.8 &    5.35 &  169.8 &  0 &  0 \\
00047+3416 &    5.6 &   2.50 &  545.0 &   0.670 &  0.623 &  136.4 &   99.4 &   95.30 &  238.0 &  0 &  0 \\
00057+4549 &   88.3 &   0.63 &  509.6 &   0.220 &  6.210 &   13.5 &   54.9 &  328.00 &  254.0 &  0 &  0 \\
00084+2905 &   33.6 &   3.37 &    0.3 &   0.535 &  0.024 &  284.4 &  105.6 &    6.70 &    5.0 &  0 &  0 \\
00093+2517 &   19.5 &   1.13 &    1.3 &   0.220 &  0.025 &  169.0 &   74.0 &   29.50 &  237.0 &  0 &  0 \\
00134+2659 &    7.0 &   2.42 &  422.0 &   0.720 &  0.641 &  193.0 &  124.1 &   18.01 &  223.7 &  1 & $-$1 \\
00174+0853 &   15.3 &   1.25 &   35.7 &   0.002 &  0.189 &  124.1 &   95.4 &    3.94 &  234.2 & $-$1 &  1 
\enddata 
\end{deluxetable*}

Table~\ref{tab:cpmvb}  contains  the  WDS  code  of  the  system,  its
parallax $\pi_{\rm HIP}$  (in mas), the mass of  the primary component
in the inner subsystem $M_1$,  five elements of the inner visual orbit
$P_1,a_1,e_1,\Omega_1,i_1$,  separation   $\rho$  and  position  angle
$\theta$ of the  outer companion, its direction of  the angular motion
(1 for direct,  $-$1 for retrograde, 0 for  unknown), and the apparent
rotation sense  (1 for  co-rotating, $-1$ for  counter-rotating).  The
sense of angular  motion is considered known if  the position angle of
the outer component, corrected for  precession, has changed by no less
than 2\degr ~between  its first and last observations  reported in the
Washington Double  Star Catalog,  WDS \citep{WDS}. However,  the first
measures in the WDS are  not always accurate, especially those made in
the  18th  century.  Some  genuine  tertiary  companions confirmed  by
common radial velocity or by the large common proper motion may appear
to  move  faster  than  the  escape  velocity  because  of  inaccurate
positions.   We consider  the first-epoch  measurements of  those wide
pairs  as unreliable  and assume  that their  revolution  direction is
unknown.   Table~\ref{tab:cpmvb} contains  443 systems,  of  which 216
have known revolution direction of  the outer pair,  including six
  2+2 systems with both inner orbits known.

For the sign correlation, there is  no need to know the inner orbit, only
the sense of  rotation. About a hundred additional  triples with known sense
of inner and outer rotation can be found in the MSC. Such increment of
the sample size  is not critical, so we decided  not to consider those
additional objects.  Inner orbital  elements are useful for comparison
of   eccentricities  and   for  the   anlysis  of   configurations  in
Section~\ref{sec:config}.

Orbital  motion  in the  inner  subsystem  introduces some  modulation
(wobble) in the angular motion  of the outer pair.  The average effect
of this modulation is zero,  so it is irrelevant when the observations
cover more than one inner period.  We neglect this complication, which
might distort the measured direction of motion in a few systems, if at
all. When  accurate {\it  instantaneous} proper motions  are available
from {\it Gaia}, their distortion  by the inner subsystems will need to
be addressed carefully.

The sample of triple systems used here is based on the compilative MSC
catalog affected by various poorly  known selection biases.  It is not
representative of  the real population of hierarchical  systems in the
field.  Additional selection  effects are  caused by  the need  to get
measurable  revolution  direction of  the  outer  system (this  favors
smaller outer separations).  Yet,  the selection filters do not depend
on  the {\it  sense} of  rotation.  Similarly,  the  measured rotation
direction and eccentricity of the inner orbit should not depend on the
rotation direction of the outer  orbit.  So, the results of this study
are reasonably  immune to the selection  and are valid  for the actual
population of hierarchical systems.

\begin{figure}[ht]
\plotone{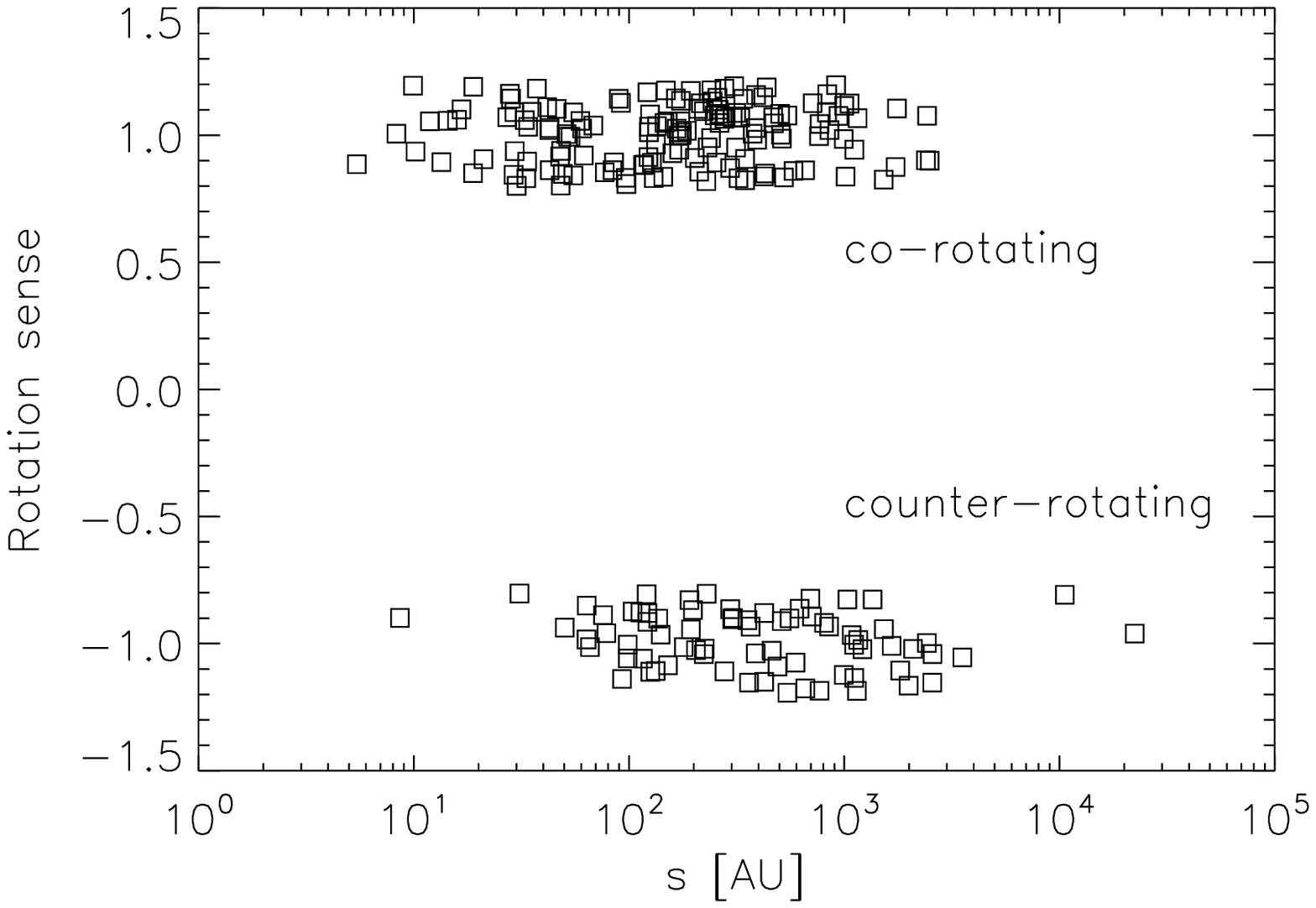}
\plotone{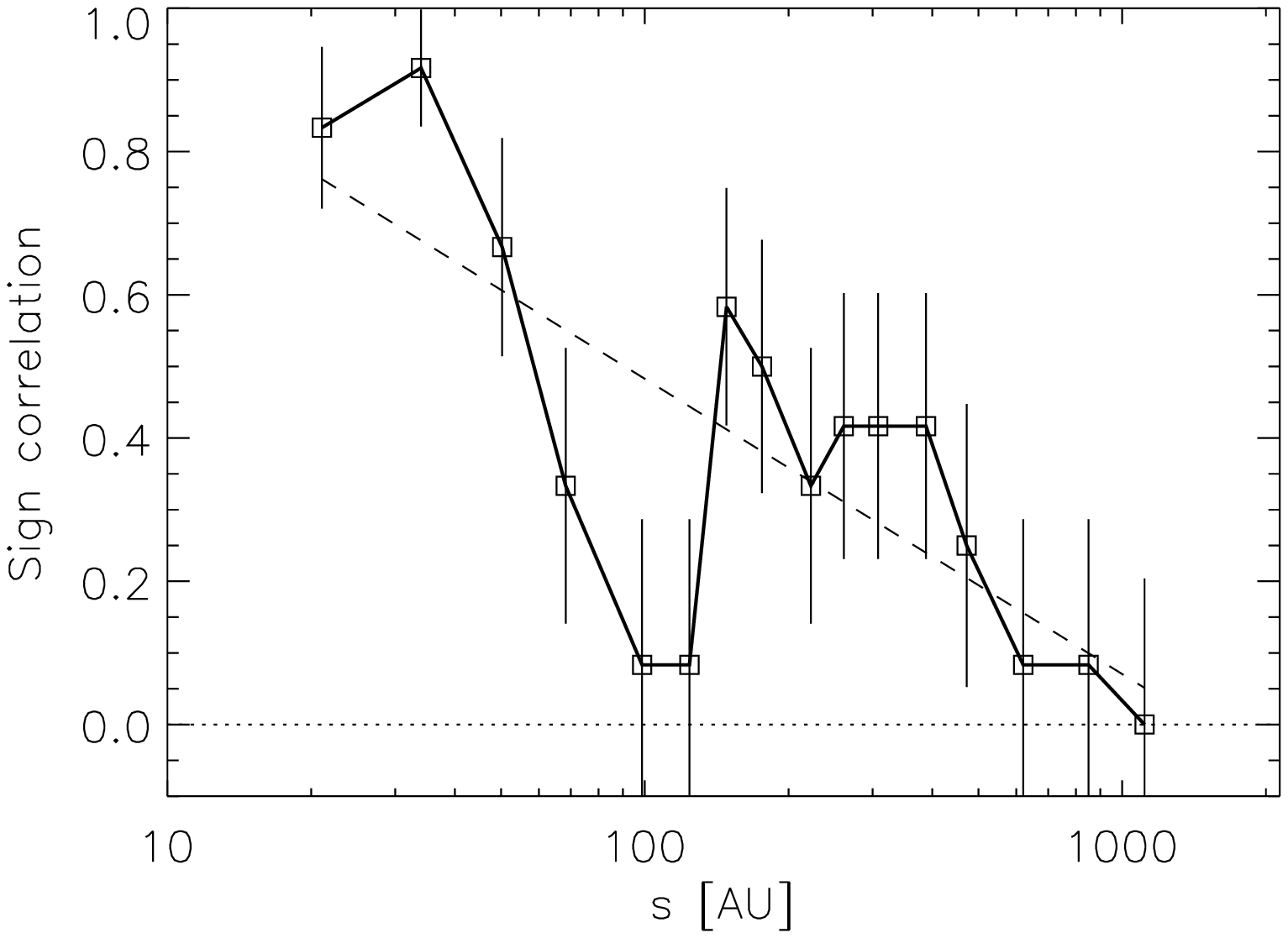}
\caption{Sign  correlation $C$  vs.  outer  projected  separation $s$.
  Top:  relative  revolution  sense   (1  for  co-rotating,  $-1$  for
  counter-rotating)  vs.  projected  separation $s$   (random vertical
  spread is introduced  to avoid overlap). Bottom: average  $C$ and its
  error as a  function of separation, computed in sub-samples of 24
  systems  as  a  running   mean  and  plotted  against  the  median
  separation in each sub-sample. The dashed line is a linear fit.
\label{fig:cum} }
\end{figure}

\section{Sign correlation}
\label{sec:sign}

\subsection{Definition}

The sign correlation $C$ is based on counting the number of triples
with coincident revolution directions $n_+$ and the number of
apparently counter-rotating triples $n_-$. Then 
\begin{equation}
C = (n_+ - n_-)/(n_+ + n_-) = 2 n_+/n - 1 .
\label{eq:C}
\end{equation}
  It  can  be  inferred   from  the  properties  of  the  binomial
  distribution that  the rms  error of this  estimate is  $\sigma_C =
\sqrt{ (1 +C) (1-C)/n}$, where $n = n_+ + n_-$ is the sample size.  It
is  easy to show  that for  random orbit  orientation relative  to the
observer, the average angle between angular momentum vectors $\Phi$ is
related   to   $C$,   $\langle   \Phi   \rangle  =   \pi/2(1   -   C)$
\citep{Worley1967}.  The revolution  direction of the inner subsystems
is securely measured from the  inclination $i$ of their orbits: direct
if  $ i  < 90\degr$,  retrograde otherwise.   The full  sample  of 216
triples  yields $C  = 0.324  \pm 0.064$,  or $\langle  \Phi  \rangle =
60\fdg8$, in agreement with \citet{ST02} and earlier works.

The relation  between sign correlation  and $\langle \Phi  \rangle$ is
valid for  random orientation  with respect to  the observer.   In the
present sample  is is  not quite random  for two reasons.   First, the
computation of   visual orbits  is difficult for  large inclinations
and such orbits are under-represented in VB6, despite the fact that $i
=  90\degr$ is  the  most probable  inclination  of randomly  oriented
orbits.  Second, the projected angular velocity of the outer component
also    depends   on   the    inclination,   favoring    smaller   $i$
\citep[see][]{ST02}. However,  all biases are  symmetric with respect
to the  revolution direction,  so the parameter  $C$ is a  very robust
diagnostic of the relative orbit alignment.

\subsection{Dependence of  orbit alignment on separation}
\label{sec:sep}

It has been noted by  \citet{ST02} that the orbit alignment depends on
the degree  of hierarchy, being  stronger for systems  with comparable
inner and outer  periods or separations.  This result  is confirmed by
the  new,  larger  sample.   An  even  stronger  dependence  of  orbit
alignment on the projected outer separation $s = \rho/\pi_{\rm HIP}$ is
found  here.   The  sample  has  been  sorted  on  $s$  and  the  sign
correlation $C$ was computed for  groups of 24 triples with increasing
separation,  as  a   running  mean.   Figure~\ref{fig:cum}  shows  the
dependence of the sign correlation on the outer separation. The linear
fit $C  = 1.31  - 0.41 \log  s$ is  an adequate representation  of the
trend.  The  local  minimum at  $s  \sim  100$  AU  is most  likely  a
statistical fluctuation.   Relatively tight triples  with $s <  50$ AU
are strongly aligned, with $C$  exceeding 0.8 or $\langle \Phi \rangle
< 18\degr$.  The top panel  of Figure~\ref{fig:cum} shows the raw data
without any binning in separation.

\subsection{Dependence of  orbit alignment on mass}
\label{sec:mass}

\begin{figure}[ht]
\plotone{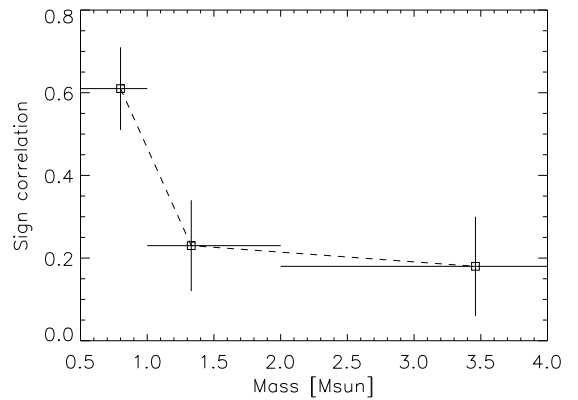}
\caption{Dependence of the sign  correlation $C$ on the mass. The vertical
  bars  depict formal  errors of  $C$, the horizontal  bars show  the mass
  range of each group.
\label{fig:cormass} }
\end{figure}

The multiplicity  fraction and  companion fraction strongly  depend on
stellar mass, being larger for massive stars. The orbit alignment also
depends on mass, but in the opposite sense, with low-mass stars having
stronger  alignment.   The  sample  has  been  subdivided  into  three
approximately  equal parts  based on  the  primary mass  in the  inner
subsystem $M_1$  (when the primary  component is itself a  binary, its
total  mass is considered).   Table~\ref{tab:mass} indicates  that the
orbit  alignment decreases with  mass.  Its  columns contain  the mass
range, the  median mass, the number  of systems $N$, the  sign correlation
$C$ and its error, median  outer separation, and average inner orbital
eccentricities     for    co-     and     counter-rotating    systems.
Figure~\ref{fig:cormass} shows the mass dependence graphically.

\begin{deluxetable}{cc cc ccc}
\tabletypesize{\scriptsize}
\tablewidth{0pt}
\tablecaption{Dependence of orbit alignment on mass \label{tab:mass}}
\tablehead{\colhead{$M_1$} &
\colhead{$\langle M_1 \rangle$} &
\colhead{$N$} &
\colhead{$C$} &
\colhead{ $\langle s \rangle$} &
\colhead{$e_+$} &
\colhead{ $e_-$ } \\
\colhead{(${\cal M}_\odot$)} & \colhead{(${\cal M}_\odot$)} & & &
\colhead{(AU)} & & 
}
\startdata
$<$1     & 0.80 & 62    & 0.61 $\pm$0.10 & 128 & 0.36 & 0.61 \\
1 to 2   & 1.33 & 83    & 0.23 $\pm$0.11 & 255 & 0.40 & 0.47 \\
$>$2     & 3.46 & 71    & 0.18 $\pm$0.12 & 315 & 0.50 & 0.47 \\ 
All      & 1.39 & 216   & 0.32 $\pm$0.06 & 222 & 0.42 & 0.49    
\enddata
\end{deluxetable}

\begin{figure}[ht]
\plotone{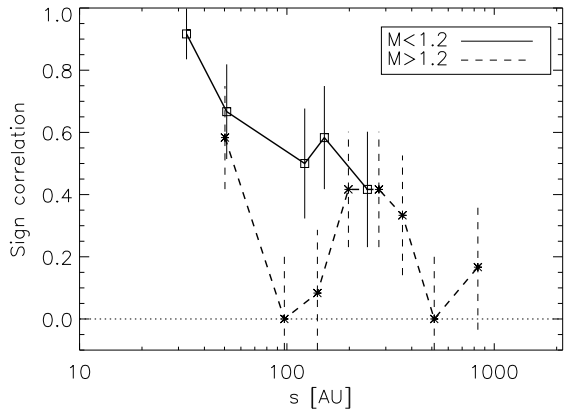}
\caption{Sign correlation  vs.   projected
  separation $s$.   The full line  corresponds to  88
  systems with $M_1  < 1.2 M_\odot$, the dashed line corresponds to
  128 systems with $M_1 > 1.2 M_\odot$.
\label{fig:cum2} }
\end{figure}

The separations of low-mass  triples are, on average, smaller compared
to the more  massive ones. Given the dependence  of orbit alignment on
the outer separation, one might  wonder whether the mass dependence is
not    caused    only    by    the    difference    in    separations.
Figure~\ref{fig:cum2} shows the dependence of orbital alignment on the
outer separation  in the  two mass regimes.   The different  degree of
orbit alignment {\it at comparable outer separation} tells us that the
mass dependence is genuine.

The  last  two  columns   of  Table~\ref{tab:mass}  contain  the  mean
eccentricity  of  the  inner  orbits  computed for  the  triples  with
coincident   and  opposite   sense  of   rotation,  $e_+$   and  $e_-$
respectively. When there  is an orbit alignment (large $C$),  we find that
$e_+ < e_-$, meaning that the inner  orbits in aligned triples are, on
average, less eccentric.

\section{Triple systems with two known orbits}
\label{sec:vbvb}

\begin{deluxetable*}{l c  rrrrr  c l r }    
\tabletypesize{\scriptsize}  
\tablecaption{Hierarchical systems with two known visual orbits (fragment)  
\label{tab:vbvb} }   
\tablewidth{0pt}   
\tablehead{                        
\colhead{WDS} & \colhead{In/Out} & \colhead{$P$} & \colhead{$e$} & \colhead{$a$} & \colhead{$\Omega$ } & \colhead{$i$} & \colhead{Orbit} & 
\colhead{Reference} & \colhead{$\Phi$} \\ 
 &     & \colhead{(yr)} &  &  \colhead{(\arcsec)} & \colhead{(\degr)}
&  \colhead{(\degr)} & \colhead{Grade} & \colhead{Code} &  \colhead{(\degr)}
   }   
\startdata     
00247$-$2653 & I &  17.25 &   0.017 &  0.460 &   14.8 &   62.0 & 2 & Tok2017b&    1.0 \\
00247$-$2653 & O &  77.47 &   0.026 &  1.531 &   13.9 &   62.6 & 4 & Tok2017b&  124.6 \\
00321+6715   & I &  15.00 &   0.083 &  0.348 &  175.0 &   47.0 & 9 & Ana2011 &    0.3 \\
00321+6715   & O & 222.30 &   0.293 &  3.322 &  174.9 &   47.3 & 5 & Doc2008d&   94.3 \\
00335+4006   & I &   4.72 &   0.076 &  0.058 &   96.1 &   97.1 & 4 & Tok2017a&  142.6 \\
00335+4006   & O &  69.37 &   0.329 &  0.389 &  299.0 &  112.8 & 2 & Tok2017a&   27.1 \\
00568+6022   & I &   4.85 &   0.224 &  0.032 &  149.9 &   47.6 & 4 & Doc2006c&   20.8 \\
00568+6022   & O &  83.10 &   0.241 &  0.245 &  175.0 &   54.9 & 2 & CWA1992 &   99.2 
\enddata 
\end{deluxetable*}

The  sign  correlation  constrains  the average  angle  $\langle  \Phi
\rangle$,  but not  its  distribution. A  mixure  of well-aligned  and
randomly  aligned systems or  a single  population of  loosely aligned
systems  can  have  the   same  $\langle  \Phi  \rangle$.   Additional
information on the distribution of  $\Phi$ can be obtained from triple
stars with known inner and outer orbits studied in this  Section and from the
apparent      configurations      of      triples      studied      in
Section~\ref{sec:config}.

The  angle  $\Phi$ between  the  angular  momentum vectors,  sometimes
called  mutual inclination, is computed as
\begin{equation}
\cos \Phi = \cos i_1 \cos i_2 + \sin i_1 \sin i_2 \cos(\Omega_1 - \Omega_2) ,
\label{eq:phi}
\end{equation}
where $i$ and $\Omega$ are the inclinations and position angles of the
node in the inner and outer orbits.

Visual  orbits  do not  distinguish  between  the  two orbital  nodes,
leaving  a  $180\degr$  ambiguity   in  the  element  $\Omega$.   To
determine  the angle  $\Phi$, we  thus  need to  identify the  correct
ascending nodes in both inner and outer orbits from radial velocities.
This is  done only for  a small number  of systems.  The  ambiguity in
$\Omega$ is  equivalent to   the $\pm$ sign  of the second  term in
equation~\ref{eq:phi}.  So,  for each  system the two  angles $\Phi_1$
and $\Phi_2$  are computed and  we do not  know which of those  is the
correct one.   The work  around consists in  simulating the  effect of
this ambiguity numerically.

In the  above sample, 54 triples  have known outer  visual orbits. The
actual number is larger, but  we discarded poorly defined outer orbits
with very long periods above  1000 yr and inner orbits with incomplete
parameters (e.g. the missing angle  $\Omega$).  Many orbits in the VB6
are  preliminary  or uncertain,  being  based  on incomplete  coverage
and/or  on  noisy  position  measurements. However,  discarding  those
orbits would  dramatically reduce the sample size.  The uncertainty of
$\Phi$ computed  from the  visual orbits is  difficult to  quantify in
most cases.

Table~\ref{tab:vbvb}  lists  relevant data  for  this sub-sample,  two
lines per system.  The first  line gives the orbital parameters of the
inner binary  ('I' in the  2nd column): its period $P$,  eccentricity $e$,
semimajor axis  $a$, position angle  of node $\Omega$  and inclination
$i$, the grade of  the orbit, from 1 (best) to 5  (tentative), 8 and 9
(astrometric),  and the  bibliographic reference  code adopted  in the
VB6.  The last column gives  the angle $\Phi_1$.  The following O-line
contains the orbital  parameters of the outer binary,  while the angle
$\Phi_2$   is   given   in   the   last  column.    Some   orbits   in
Table~\ref{tab:vbvb} were determined or  refined by the author and are
still unpublished. 

For this sample, we find $C = 0.48 \pm 0.12$, or $\langle \Phi \rangle
= 47\degr$.   These systems  are closer than  in the main  sample; the
average outer semimajor axis is only 75\,AU.

\begin{figure}[ht]
\plotone{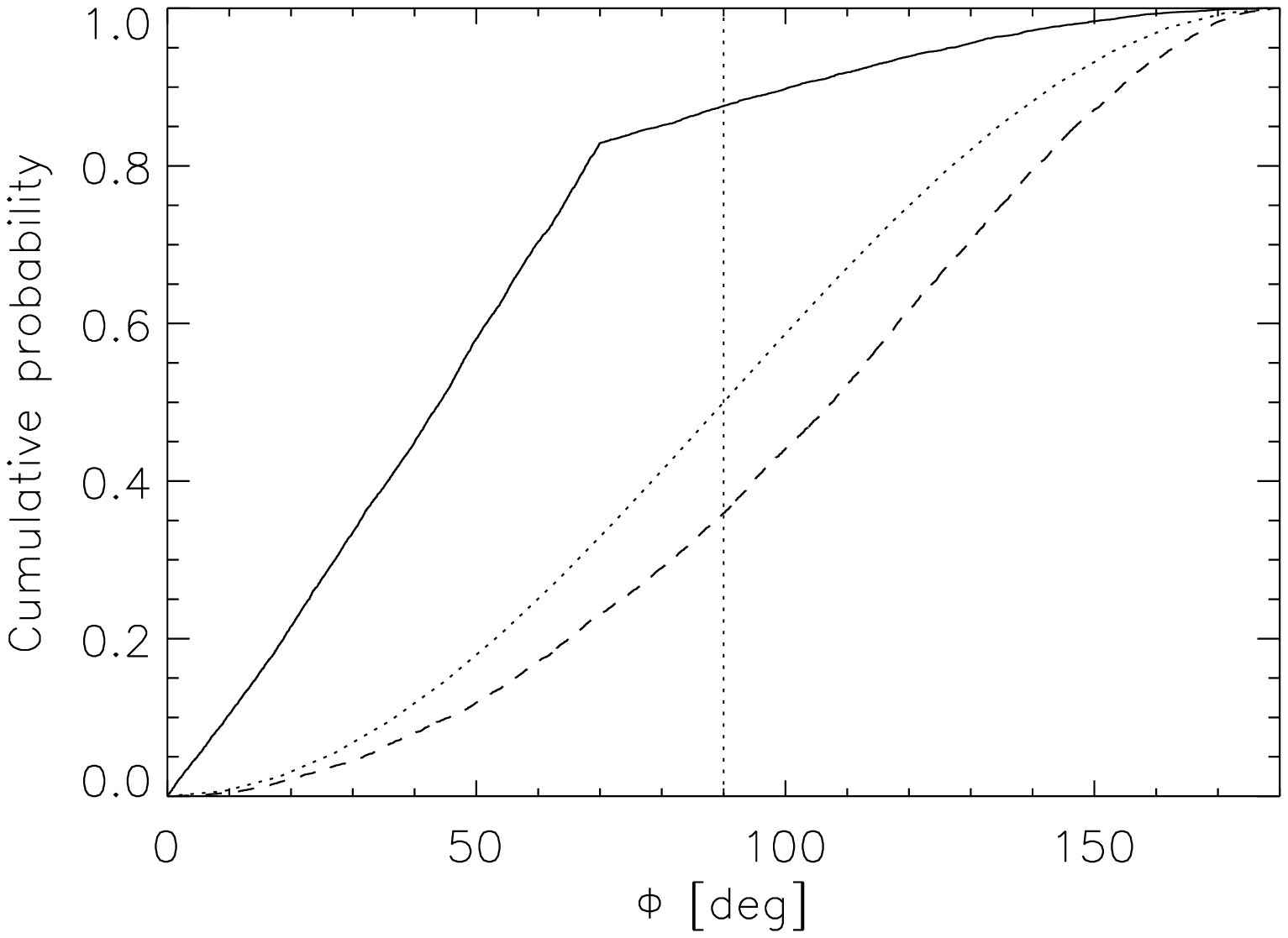}
\plotone{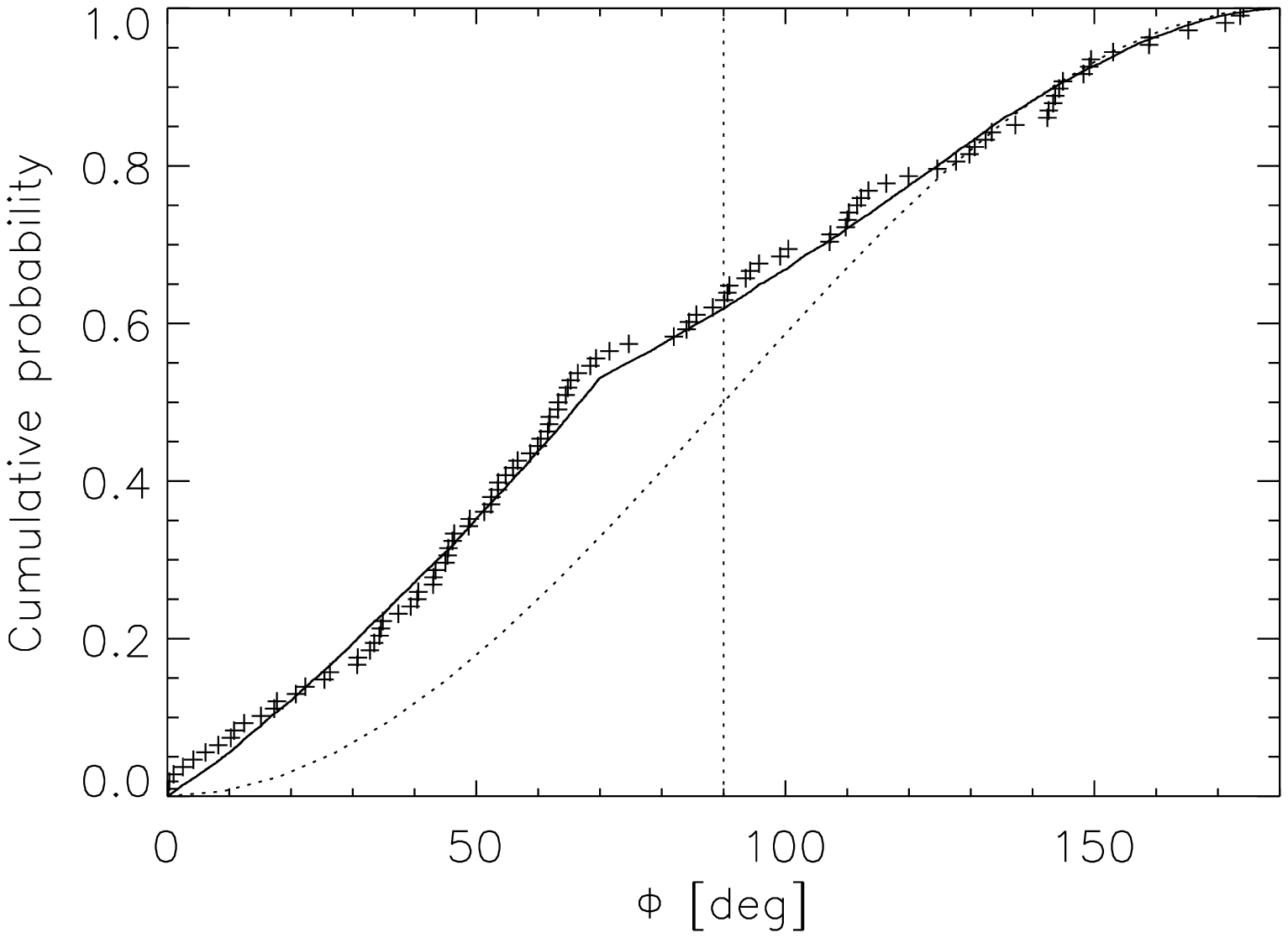}
\caption{Top:  cumulative  distribution  of  the angles  $\Phi_1$  and
  $\Phi_2$ (full  and dashed lines, respectively)  in simulated triple
  stars. Bottom:  joint cumulative distributions of both  angles in the
  real  sample (crosses)  and  in simulations  (full  line).  In  both
  plots,  the  dotted  line  corresponds  to  the  uncorrelated  orbit
  orientation.
\label{fig:vbvb} }
\end{figure}

When the  orbits are  aligned randomly, both  $\cos \Phi_1$  and $\cos
\Phi_2$ are  distributed uniformly, and these two  angles are slightly
anti-correlated with each other.  When the orbital spins are partially
aligned,  the cumulative distribution  of the  true angle  $\Phi_1$ is
above the random one, while the cumulative distribution of $\Phi_2$ is
below.  The statement  of \citet{ST02}  that for  partially correlated
spins the distribution of $\Phi_2$ is random is incorrect.

We simulated triple systems with some assumed distribution of $\Phi$ and
random orientation  with respect to  the observer.  A simple  model is
adopted, where  $\Phi$ is distributed  uniformly in the  interval $(0,
\Phi_0)$  for a  fraction $f$  of the  triples and  is random  for the
remaining $1-f$ fraction. After a few trials, the parameters $\Phi_0 =
70\degr$ and $f=0.8$ were chosen to match the combined distribution of
both angles in the  real sample (Figure~\ref{fig:vbvb}, bottom).  This
model has $\langle  \Phi \rangle = 46\degr$.  The  top panel shows the
separate  cumulative distributions  of  $\Phi_1$ and  $\Phi_2$ in  the
simulated  triples.  The  distribution  of $\Phi_1$,  as  well as  the
merged distribution,  have a characteristic break at  $\Phi \approx \Phi_0$,
and the  merged cumulative distribution  of $\Phi$ is almost  (but not
exactly)  linear  at  $\Phi  <  \Phi_0$,  reflecting  the  true  input
distribution  of  $\Phi_1$.   A  population  of  well-aligned  (nearly
coplanar) triples would  manifest itself by a peak at  small $\Phi$  in
the merged distribution, which  is not present. However, the empirical
distribution of  $\Phi$ is  broadened by the  errors of  visual orbits.

We  see that  despite the  uncertainty associated  with  unknown orbit
nodes,  the  merged  distribution   of  both  angles  $\Phi$  contains
information  on the  distribution of  the true  angles  $\Phi_1$.  The
observed distribution  matches the model where 80\%  of visual triples
are aligned within $70\degr$.
%

\section{Projected configurations}
\label{sec:config}

Another method of checking orbital  alignment in triple stars is based
on their apparent (projected) configurations.  It does not require the
knowledge  of rotation  sense and  can be  applied even  to  very wide
triples.  Suppose that a coplanar  triple system is seen edge-on. Then
the position angles of the inner  and outer pairs will be either equal
or will differ by $180\degr$.  When such system is viewed from an arbitrary
direction and at   arbitrary phases of both  orbits, the correlation
between  the   position  angles  is  reduced,  but   does  not  vanish
completely.  Our simulations show that orbit alignment can be detected
from apparent configurations in a large (on the order of 1000 or more)
sample of  triple stars with strong coplanarity,  otherwise the effect
is washed  out by  the randomness of  projections and  orbital phases.
Therefore, we do not study here the difference between position angles
of the inner and outer pairs.

\begin{figure}[ht]
\centerline{
\plotone{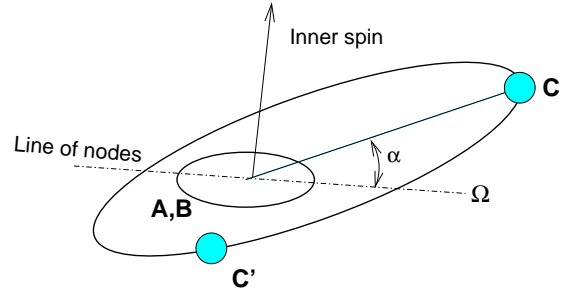}
}
\caption{Definition of  the angle $\alpha$  between the inner  line of
  nodes in  the subsystem A,B and  the position angle  of the tertiary
  component C.
\label{fig:config} }
\end{figure}

A better option  is to compare the angle $\alpha$  between the line of
nodes  in  the inner  orbit  and  the  direction toward  the  tertiary
component  (Figure~\ref{fig:config}).    This  approach  requires  the
knowledge  of  the  inner  orbit,  but eliminates  one  random  factor
(position of  the inner binary on  its orbit).  The  angle $\alpha$ is
defined modulo  90\degr.  The  line of nodes  is perpendicular  to the
projection of  the orbital  spin vector on  the sky.  As  the tertiary
component  C spends  more time  at  large separations,  the chance  of
obtaining  small  $\alpha$  for  a  nearly coplanar  triple  is  high.
However, when the tertiary is  found at another position C', the angle
$\alpha$ can be close to 90\degr ~even in a perfectly coplanar triple.
The method works  only in the statistical sense,  through the analysis
of  the distribution  of $\alpha$.   This  approach has  been used  by
\citet{Wheelwright2011}  to  study alignment  between  dust disk  and
orbital plane in 20 young wide binaries.

In  the pre-computer  era,  \citet{Agekyan} proposed  this method  for
evaluating   coplanarity   of  triple   stars   from  their   apparent
configurations.   He   demonstrated  analytically  that   the  average
quantity $A = \langle \cos^2 \alpha \rangle$ equals 0.6932 in the case
of coplanar orbits,  while it is 0.5 for  uncorrelated orbital planes.
In  other words,  small values  of $\alpha$  dominate in  the coplanar
case, but for the  random relative orientation $\alpha$ is distributed
uniformly between  0 and  90\degr.  Neither Agekyan  nor \citet{Tok93}
had  detected any  significant deviations  of  $A$ from  0.5 in  small
samples of  visual triple  stars.  Here we  study the  distribution of
$\alpha$ rather than the average diagnostic $A$.

\begin{figure}[ht]
\plotone{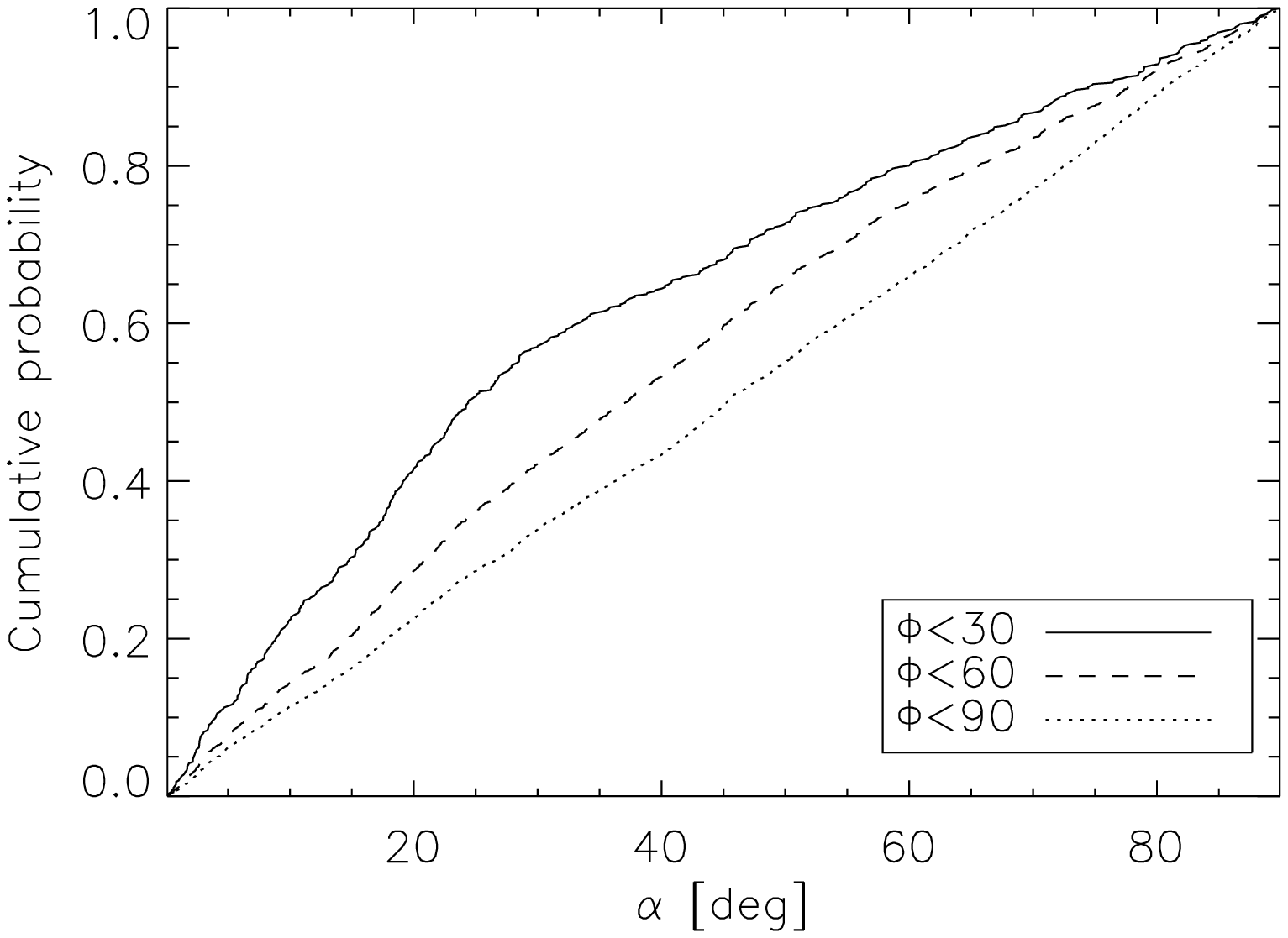}
\plotone{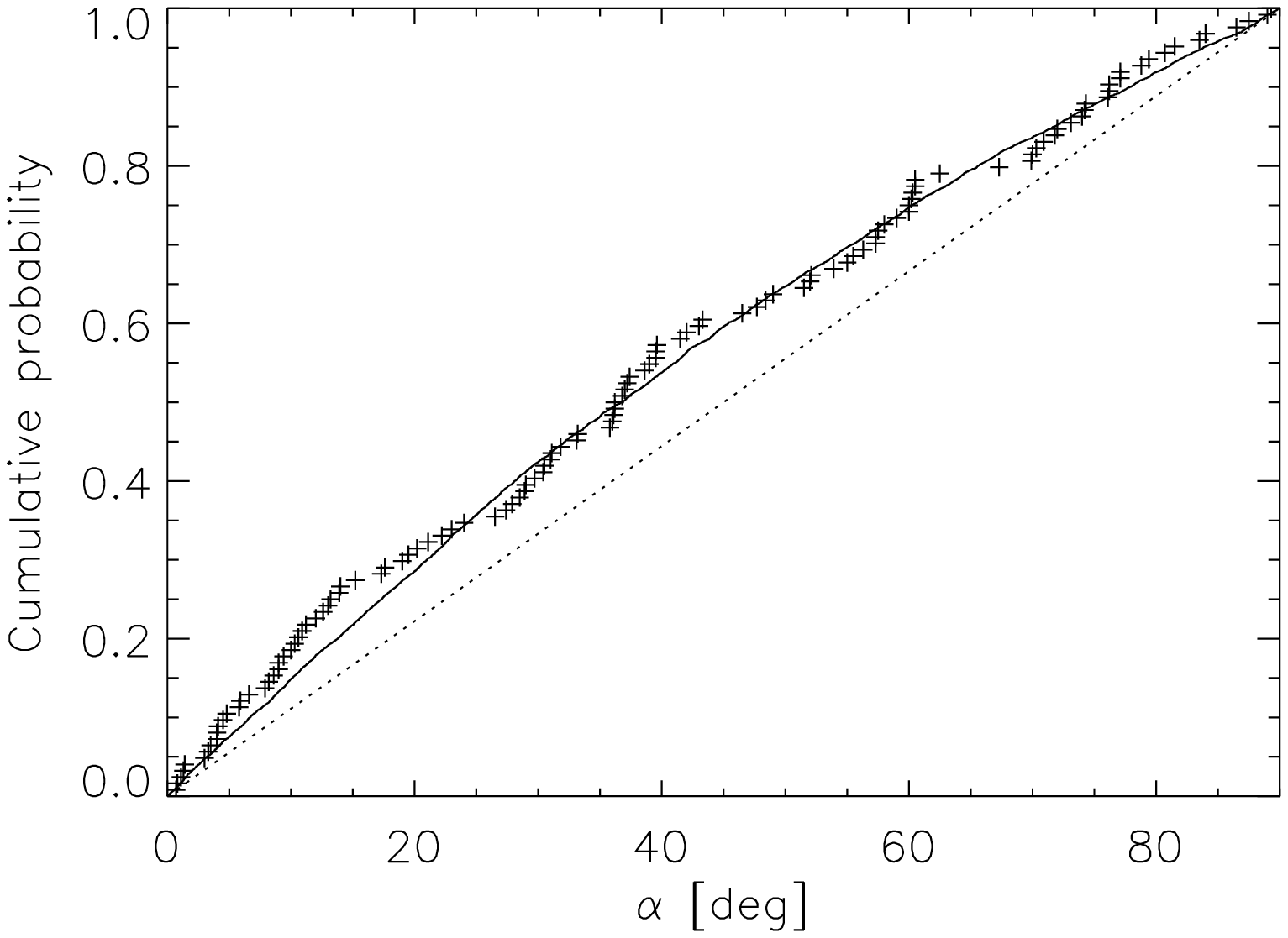}
\caption{Cumulative  distributions of the  angle $\alpha$.   The upper
  plot  shows the results  of numerical  simulation.  The  bottom plot
  shows the actual distribution in the subset of 124 systems with $M_1
  <  1  {\cal  M}_\odot$   (crosses)  compared  to  the  simulated
  distribution  (full  line).   The  dotted  line  marks  the  uniform
  distribution corresponding to the uncorrelated orbit orientation.
\label{fig:alpha} }
\end{figure}

The distribution  of $\alpha$ was calculated  by numerical simulation.
A large number of triple  systems with random relative orientation have
been generated.   The distribution of $\alpha$ for  the full simulated
sample is  uniform, as  expected.  When only  the aligned part  of the
sample with  $\Phi < \Phi_0$  is selected, small $\alpha$  become more
probable    and   the    distribution    deviates   from    uniformity
(Figure~\ref{fig:alpha}, top).  For three values of $\Phi_0 = 30\degr, 60\degr,
90\degr$, the  average angle  between the  orbital spins is  $\langle \Phi
\rangle = 20\fdg4, 39\fdg5, 57\fdg1$.  With $\Phi_0 =90\degr$, the average
$\Phi$ is similar to the  one actually observed, the orbital spins are
partially  correlated.   However,  the  distribution  of  $\alpha$  is
practically indistinguishable  from the uniform one.   So, this method
can detect only a relatively  well-aligned population of multiples. Even
in this  case (e.g.  $\Phi  < 30\degr$), the  $\alpha$-distribution is
strongly broadened  by random projection and  orbital phase \citep[see
  also Figures 2 and 3 in][]{Wheelwright2011}.

The angles  $\alpha$ were computed for  all 443 triples  in our sample
(there  is no  need  to know  the  revolution direction  of the  outer
component). Their  distribution only  barely differs from  the uniform
one.   This difference  is enhanced  in  the sub-sample  of 140  close
triples  with outer separation  $s <  300$ AU,  in agreement  with the
previous finding  that such triples  are better aligned (they  have $C
\approx 0.6$ and $\langle  \Phi \rangle \approx 36\degr$).  The effect
is even larger for the sub-sample  of 124 low-mass triples with $M_1 <
1 {\cal M}_\odot$.  The  maximum deviation of the cumulative histogram
of  $\alpha$ from the  linear (uniform)  distribution in  the low-mass
sample is 0.13 and  corresponds to the Kolmogorov-Smirnov significance
level   of  0.02.    Figure~\ref{fig:alpha}   compares  the   observed
distribution  of  $\alpha$ in  low-mass  triples  with simulations  of
partially  aligned triples  presented above  in Section~\ref{sec:vbvb}
($\Phi_0 =  70\degr$, $f=0.8$).   The agreement is  satisfactory, thus
favoring the simulated distribution of $\Phi$.

\section{Discussion}
\label{sec:disc}

We  found a strong  tendency of  orbit alignment  in triple  stars with
outer  separations less than  $\sim$50 AU.   This roughly  matches the
scale of circumstellar disks.  Formation and/or dynamical evolution of
those  close  triples  should   have  been  influenced  by  the  disk.
Additional  evidence  of   the  importance  of  dissipative  dynamical
interactions  with  gas  is  furnished by  the  statistically  smaller
eccentricity of  inner orbits in  the co-aligned triples,  compared to
the   average    eccentricity.    The   quadruple    system   HD~91962
\citep{Planetary} with  nearly coplanar orbits of  small eccentricity is
thus  representative  of the  class  of  low-mass  hierarchies with  a
planar,  planetary-like  architecture.   There  exist  other  low-mass
multiples with similar  properties \citep{Tok17}.

We also found that the orbit alignment is stronger in triple stars with
low-mass primaries,  compared to more  massive triples. Compact  ($s <
50$  AU) low-mass  triples have  $C  \approx 0.8$,  or $\langle  \Phi
\rangle \approx 18\degr$,  while more massive triples are less well aligned.

Dynamical  interactions in  unstable multiples  should  leave residual
misaligned    triples,   often    with    highly   eccentric    orbits
\citep{Antognini2016}. It seems that chaotic stellar dynamics played a
larger role in  the formation of massive stars.   Another process that
can create  misaligned triples is  the accretion of gas  with randomly
aligned angular momentum at the epoch of star formation. Massive stars
form in clusters  and accrete misaligned gas from  the cluster volume,
not just from the parent  core. Misaligned gas changes the orientation
of the outer orbit and, even more importantly, causes its rapid inward
migration.  Shrinking  of the outer  orbit can destabilize  a multiple
system   \citep{Smith1997},  leading   to  violent   interactions  and
ejections  of some members  at high  velocity (runaway  stars).  These
internal  dynamical  interactions operate  even  on  small scale.   In
contrast,  dynamical  interactions  with  other  cluster  members  are
relevant on  the spatial  scale of thousands  of AU (depending  on the
cluster density) and are associated with moderate ejection velocities.

Alignment in triple systems is  related to the alignment between disks
and stellar spins in binaries, being influenced by the same phenomena.
\citet{Monin2006} estimated  from polarization the  relative alignment
between two disks in young wide binaries.  They found a clear evidence
of disk  alignment between  binary components and  a hint  on stronger
alignment    in    closer     binaries.     Echoing    this    result,
\citet{Wheelwright2011}  established that  resolved disks  are aligned
with the  binary orbit.  Recently,  angles between projected  spins of
young stars  (traced by  the outflow direction)  paired in wide  ($s >
1000$  AU)   binaries  have   been  studied  by   \citet{Lee2016}  and
\citet{Offner2016}.\footnote{Misleadingly,    these    authors    call
  orthogonal  directions ``anti-aligned'',  while  here and  generally
  anti-alignment  means  oppositely  directed parallel  spins.}   This
technique is similar to  the $\alpha$-statistics for triple stars, but
it eliminates one random factor  (the phase of the outer orbit), being
affected only by  random projections.  Both triple-star configurations
and the distribution of projected spin angles are not sensitive to the
spin direction.  This is the  weakness of this method, compared to the
sign  correlation.  Unlike Monin  et al.,  \citet{Offner2016} conclude
that  the spin  directions  of  components in  wide  binaries are  not
mutually correlated.   Their sample is  small (26 spin pairs)  and the
significance of this result is  marginal.  It matches however the lack
of alignment  between outer  and inner orbits  in triple  systems with
outer  separations above  1000\,AU, found  here.

On the other hand, orbit  alignment is strong in very compact multiple
systems.  \citet{Borkovits2016} studied  relative orbit orientation in
close triple systems containing eclipsing binaries, using {\it Kepler}
photometry  in combination  with dynamical  analysis.  This  method is
indirect because  the systems are  not spatially resolved.   The outer
periods of those  triples are of the order of a  year, their masses of
the order of one solar mass.   For 62 systems the authors were able to
estimate the  angles $\Phi$.  Borkotits  (2017, private communication)
cautioned however that the sign of $\cos \Phi$ depends on the higher order
perturbation terms and is determined by this method less reliably than
$\sin   \Phi$.     Only   one    triple out of 62  has   $\Phi    =   147\degr$
(counter-rotating),  all remaining  triples have  $\Phi<60\degr$.  The
average  $\langle \Phi  \rangle =  21\degr$ computed  from  their data
implies $C=0.77$  and matches the large  values of $C$  found here for
compact triples.

Figure~15 of  \citet{Borkovits2016} presents the  bimodal distribution
of  the  angle  $\Phi$, with  a  strong  peak  of 29  nearly  coplanar
($\Phi<10\degr$)  triples  and  the   second  peak  at  $\Phi  \approx
40\degr$,  presumably matching  the  outcome of  the Kozai  mechanism.
These  authors note,  however,  that the  inner  periods of  eclipsing
binaries  show  neither correlation  with  $\Phi$  nor the  clustering
between 3 and  10 days expected from the  tidal circularization.  This
contradicts the predictions  of \citet{Fabrycky2007}, unless the inner
periods were  shortened by other mechanisms such  as magnetic braking.
If some co-rotating inner  binaries with $\Phi \approx 40\degr$ indeed
resulted  from  the Kozai  cycles  with  tidal circularization,  their
progenitors  had  $\Phi<90\degr$,  i.e.   had correlated  rather  than
random orbital spins.  Overall, 43/62=0.69 fraction of compact triples
in \citet{Borkovits2016} have $\Phi<30\degr$, too small to result from
the Kozai  mechanism.  Instead,  gas friction in  a disk could  be the
dominant  formation channel  of  close binaries  and compact  triples.

 The spin-orbit angle and the sense of rotation can be established
  for  transiting  exoplanets, informing  us  on  the primordial  disk
  alignment.   \citet{Fielding2015} found  from  their simulations  of
  turbulent  fragmentation   typical  spin-orbit  angles   $\Phi  \sim
  40\degr$, in  agreement with observations of  exoplanets and similar
  to the orbit alignment in close triple systems.

Although the sample of triple  stars with known inner orbits and known
sense   of  revolution   of   the  third   components  has   increased
substantially since  the work  of \citet{ST02}, from  135 to  216, the
progress remains slow, being paced by the long orbital periods and the
slow  accumulation of data.  Continued monitoring  of triple  stars by
means of speckle interferometry is needed to determine more orbits and
to   reach   closer,    more   interesting   triples.    Long-baseline
interferometry  has  begun  to  make  its contribution  in  this  area
\citep[e.g.][]{Schaefer2016}  and will  hopefully continue  doing so,
especially if fainter stars  can be observed by interferometers.  {\it
  Gaia}  might determine  astrometric  orbits in  inner subsystems  of
resolved binaries by very  accurate monitoring of positions during its
5-year  mission.   Complementary accurate  radial  velocities will  be
needed to strengthen astrometric and  visual orbits and to resolve the
ambiguity of their nodes.

\acknowledgements This work would  have been impossile without the WDS
and VB6  databases, both  maintained at  the USNO by  the team  led by
B.~Mason and W.~Hartkopf.  Observations of visual binaries and triples
have been  made during more  than two centuries  by a large  number of
astronomers,  to whom  we are  deeply indepted.   Comments  by the
  Referee are gratefully acknowledged.



\end{document}